\documentclass[journal]{IEEEtran}

\usepackage[utf8]{inputenc}
\usepackage[T1]{fontenc}
\usepackage{url}
\usepackage{ifthen}
\usepackage{cite}
\usepackage[cmex10]{amsmath}
\usepackage{graphicx}
\usepackage{amssymb} % For symbols like checkmarks if needed
\usepackage{booktabs} % For better tables
\usepackage{multirow} % For multirow cells in tables
\usepackage{xcolor} % For comments or highlighting if needed during drafting
\usepackage{flushend} % For balancing columns on the last page

\usepackage{geometry}
\begin{document}

\title{KAITIAN: Communication Framework for Efficient Collaboration Across Heterogeneous Accelerators in Embodied AI Systems}

\author{%
    \IEEEauthorblockN{Jieke Lin\IEEEauthorrefmark{1}\IEEEauthorrefmark{2},
                      Wanyu Wang\IEEEauthorrefmark{1}\IEEEauthorrefmark{2},
                      Longxiang Yin\IEEEauthorrefmark{2},
                      and Yinhe Han\IEEEauthorrefmark{2}}\\
    \IEEEauthorblockA{\IEEEauthorrefmark{1}Hangzhou Institute for Advanced Study, UCAS}\\
    \IEEEauthorblockA{\IEEEauthorrefmark{2}Institute of Computing Technology, Chinese Academy of Sciences, Beijing, China}\\
    \IEEEauthorblockA{Emails: \{linjinke23, wangwanyu23\}@mails.ucas.ac.cn, \{yinlongxiang, yinhes\}@ict.ac.cn}\\[1.2ex]
\textit{Jieke Lin and Wanyu Wang contributed equally to this work.}\\[0.8ex]
\textit{Corresponding author: Longxiang Yin \{yinlongxiang@ict.ac.cn\} }
}

\maketitle

\begin{abstract}
Embodied Artificial Intelligence (AI) systems, such as autonomous robots and intelligent vehicles, are increasingly reliant on diverse heterogeneous accelerators (e.g., GPGPUs, NPUs, FPGAs) to meet stringent real-time processing and energy-efficiency demands. However, the proliferation of vendor-specific proprietary communication libraries creates significant interoperability barriers, hindering seamless collaboration between different accelerator types and leading to suboptimal resource utilization and performance bottlenecks in distributed AI workloads. This paper introduces \textbf{KAITIAN}, a novel distributed communication framework designed to bridge this gap. KAITIAN provides a unified abstraction layer that intelligently integrates vendor-optimized communication libraries for intra-group efficiency with general-purpose communication protocols for inter-group interoperability. Crucially, it incorporates a load-adaptive scheduling mechanism that dynamically balances computational tasks across heterogeneous devices based on their real-time performance characteristics. Implemented as an extension to PyTorch and rigorously evaluated on a testbed featuring NVIDIA GPUs and Cambricon MLUs, KAITIAN demonstrates significant improvements in resource utilization and scalability for distributed training tasks. Experimental results show that KAITIAN can accelerate training time by up to 42\% compared to baseline homogeneous systems, while incurring minimal communication overhead (2.8--4.3\%) and maintaining model accuracy. KAITIAN paves the way for more flexible and powerful heterogeneous computing in complex embodied AI applications.
\end{abstract}


\IEEEpeerreviewmaketitle

\section{Introduction}
\label{sec:introduction}
\IEEEPARstart{E}{mbodied} AI systems, ranging from autonomous drones navigating complex terrains to sophisticated robotic assistants interacting with the physical world, represent a paradigm shift in artificial intelligence~\cite{Sun:DaDu-E}. These systems must perceive their environment, make intelligent decisions, and act upon them, often under strict real-time constraints and limited power budgets. To achieve the requisite computational power, modern embodied AI platforms increasingly integrate a diverse array of specialized hardware accelerators, including General-Purpose Graphics Processing Units (GPGPUs) for massively parallel computations, Neural Processing Units (NPUs) optimized for deep learning inference and training, and Field-Programmable Gate Arrays (FPGAs) for custom logic. This trend towards \textit{heterogeneous computing} promises significant performance and efficiency gains by matching computational tasks to the most suitable hardware~\cite{HeteroSurvey2020}.

However, realizing the full potential of heterogeneous architectures is fraught with challenges, particularly in the domain of inter-accelerator communication. Each hardware vendor typically provides its own proprietary communication library (e.g., NVIDIA's NCCL for GPUs, Cambricon's CNCL for MLUs). While these libraries are highly optimized for communication between homogeneous devices (e.g., GPU-to-GPU), they are often incompatible with libraries from other vendors. This "walled-garden" ecosystem creates significant interoperability barriers, making it exceedingly difficult to orchestrate collaborative computation across different types of accelerators within a single, unified application, such as a distributed deep learning training job. Consequently, developers are often forced to use only one type of accelerator or resort to cumbersome, inefficient manual data transfers via the host CPU, leading to underutilized hardware resources, increased communication latency, and stifled innovation in leveraging the combined strengths of diverse accelerators.

The problem is particularly acute in distributed training of large AI models, a common requirement for advanced embodied AI capabilities. Efficient scaling of training across multiple devices is paramount, but heterogeneity introduces complexities. Different accelerators may exhibit varying computational speeds and memory capacities, leading to load imbalances if tasks are distributed naively. Synchronizing model parameters and gradients across these disparate devices without incurring substantial communication overhead is a critical hurdle. Existing deep learning frameworks like PyTorch~\cite{Paszke:PyTorch} and TensorFlow~\cite{TensorFlow} primarily support a single communication backend per training job, further limiting native support for true heterogeneous distributed training.

To address these pressing challenges, we propose \textbf{KAITIAN}, a distributed communication framework designed to unlock the collaborative potential of heterogeneous accelerators in embodied AI systems. KAITIAN offers a software abstraction layer that intelligently manages communication across a diverse hardware landscape. The core contributions of this work are:
\begin{itemize}
    \item \textbf{A Hybrid Communication Architecture:} KAITIAN seamlessly integrates vendor-specific, high-performance communication libraries (e.g., NCCL, CNCL) for efficient intra-group communication within clusters of homogeneous accelerators, while employing a general-purpose communication backend (Gloo) for interoperable inter-group communication across different accelerator types, relayed via host CPU memory.
    \item \textbf{A Load-Adaptive Scheduling Mechanism:} To counteract performance disparities among heterogeneous devices, KAITIAN incorporates a dynamic load balancing strategy. It benchmarks accelerator performance and proportionally allocates data batches, ensuring that all devices contribute effectively and complete their computational workloads in a synchronized manner, thereby maximizing overall system throughput.
    \item \textbf{Seamless Integration with PyTorch:} KAITIAN is implemented as a pluggable communication backend for PyTorch, allowing researchers and developers to leverage heterogeneous resources with minimal changes to their existing distributed training scripts.
    \item \textbf{Empirical Validation:} We demonstrate KAITIAN's effectiveness through comprehensive experiments on a testbed comprising NVIDIA GPUs and Cambricon MLUs. Results show significant training speedups (up to 42\%) in heterogeneous configurations compared to homogeneous baselines, with negligible communication overhead and preserved model accuracy.
\end{itemize}

The remainder of this paper is organized as follows: Section~\ref{sec:related_work} discusses related work in heterogeneous communication and scheduling. Section~\ref{sec:framework_design} details the architecture and core mechanisms of KAITIAN. Section~\ref{sec:implementation_evaluation} describes the implementation and presents a thorough experimental evaluation. Section~\ref{sec:discussion} discusses the implications, limitations, and potential extensions of our work. Finally, Section~\ref{sec:conclusion} concludes the paper and outlines future research directions.

\section{Related Work}
\label{sec:related_work}
The challenges of communication and coordination in heterogeneous computing environments have been addressed from various perspectives. This section reviews relevant literature in communication libraries, frameworks for heterogeneous systems, and load balancing techniques.

\subsection{Communication Libraries and Frameworks}
Standardized communication interfaces like the Message Passing Interface (MPI)~\cite{MPIForum} have long been a cornerstone of high-performance computing (HPC). While MPI is versatile, it may not always be optimized for the specific memory hierarchies and interconnects of modern accelerators. Vendor-specific libraries such as NVIDIA's NCCL~\cite{NVIDIANCCL} and AMD's RCCL~\cite{AMDRCCL} provide highly optimized collective communication primitives for their respective GPU architectures. Similarly, Cambricon offers CNCL for its MLUs. These libraries achieve excellent performance for homogeneous device clusters.

For inter-vendor communication, solutions are less mature. Gloo~\cite{FacebookGloo}, developed by Facebook, is a collective communication library that supports various backends, including TCP/IP for CPU-based communication and direct GPU-to-GPU communication via NCCL or InfiniBand verbs. While Gloo can bridge different systems, its direct support for diverse accelerator types beyond GPUs is limited without custom extensions.

Several research projects have aimed to unify communication across heterogeneous devices. For instance, OpenCL~\cite{OpenCL} and SYCL~\cite{SYCL} provide programming models for heterogeneous platforms, but they focus more on computation offloading and kernel execution rather than high-level distributed training communication patterns. Frameworks like StarPU~\cite{StarPU} and Legion~\cite{Legion} offer task-based programming models that can manage dependencies and data movement in heterogeneous systems, but may require significant programming effort to integrate with existing deep learning workflows.
% \placeholder{Discuss 1-2 more specific research frameworks that attempt to solve heterogeneous communication, highlighting how KAITIAN differs or builds upon them. For example, are there frameworks that try to directly bridge NCCL and CNCL without CPU relay? What are their limitations?}

\subsection{Load Balancing in Heterogeneous Systems}
Effective load balancing is crucial for maximizing performance in heterogeneous environments where processing units can have vastly different computational capabilities. Static load balancing techniques assign tasks based on predetermined device characteristics, which can be suboptimal if performance varies dynamically. Dynamic load balancing strategies, on the other hand, adapt task allocation based on runtime conditions.

In the context of distributed deep learning, several approaches have been proposed. PipeDream~\cite{NarayananPipeDreamSOSP2019} introduced pipeline parallelism that can mitigate straggler effects in heterogeneous clusters, but focuses on a different parallelism dimension. Some works have explored dynamic batch sizing or work stealing. For example,~\cite{LiHeterogeneityAwareDLB2021} proposed heterogeneity-aware dynamic load balancing for distributed deep learning training, which involves adapting workload sizes.
% \placeholder{Compare KAITIAN's load-adaptive mechanism with the approach in Li et al. (2021) or other specific papers on dynamic load balancing for distributed training in heterogeneous settings.}

KAITIAN differentiates itself by providing a pragmatic and readily integrable solution within a popular deep learning framework (PyTorch). It focuses on a hybrid communication strategy that leverages the best of both worlds (vendor-optimized and general-purpose libraries) and combines it with a simple yet effective load-adaptive mechanism specifically tailored for synchronous data-parallel training across diverse accelerator types. While other frameworks might offer more comprehensive but complex solutions, KAITIAN aims for ease of use and immediate applicability to existing embodied AI research workflows.

\section{The KAITIAN Framework Design}
\label{sec:framework_design}
KAITIAN is architected to provide a flexible and efficient communication backbone for distributed applications running on heterogeneous accelerator clusters. Its design philosophy centers on modularity, leveraging existing optimized libraries where possible, and providing a clear abstraction for managing cross-device data exchange. 

\begin{figure}[t] % Changed from figure* to figure, and [htbp] to [t] for top placement
    \centering
    \includegraphics[width=\columnwidth]{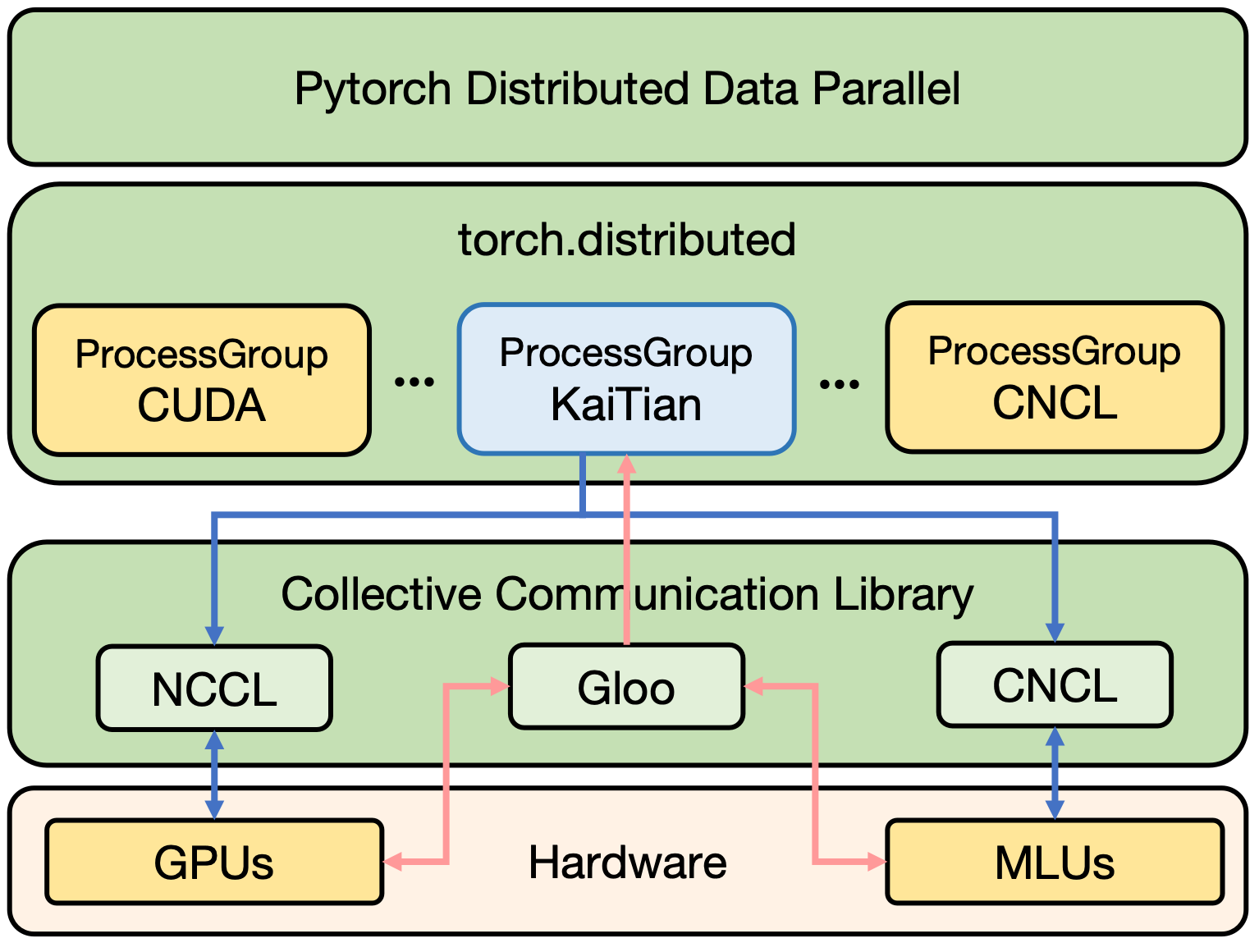} % Using \columnwidth for single column width
    \caption{KAITIAN Framework Integration with PyTorch.} % Shortened caption
    \label{fig:kaitian_architecture_single_col} % Unique label for the figure
    \vspace{-1em} % Reduced space after figure
\end{figure}

The core principle of KAITIAN is its integration into PyTorch's distributed ecosystem, as illustrated in Figure~\ref{fig:kaitian_architecture_single_col}. KAITIAN introduces a new custom ProcessGroup, `ProcessGroupKaiTian`, into `torch.distributed` using PyTorch's C++ extension mechanism. When PyTorch's DistributedDataParallel (DDP) module is initialized with this custom ProcessGroup, all subsequent collective communication calls (e.g., AllReduce for gradient synchronization, broadcast for model synchronization) are routed through KAITIAN. This allows KAITIAN to intelligently manage how these communication primitives are executed across a mix of hardware. Specifically, for operations involving only homogeneous accelerators (e.g., a group of NVIDIA GPUs), KAITIAN dispatches the calls to highly optimized vendor-specific libraries like NCCL. Similarly, for operations confined to Cambricon MLUs, CNCL is used. These intra-group homogeneous communications are depicted by blue data paths in Figure~\ref{fig:kaitian_architecture_single_col}. For inter-group communication that spans different accelerator types (e.g., aggregating gradients from both GPUs and MLUs), KAITIAN utilizes the Gloo collective communication library, where data is relayed via host CPU memory. This heterogeneous communication path is shown with pink data paths.

The primary challenge KAITIAN addresses is the inherent limitation in mainstream deep learning frameworks like PyTorch, which typically bind an entire distributed training job to a single communication backend (e.g., NCCL, Gloo, or MPI). This prevents the native, simultaneous use of multiple vendor-specific libraries optimized for different hardware. `ProcessGroupKaiTian` acts as a meta-backend or a dispatch layer to overcome this.

\subsection{Hierarchical Communication Management}
KAITIAN's `ProcessGroupKaiTian` manages communication across diverse accelerators by distinguishing between operations within groups of homogeneous accelerators and operations spanning heterogeneous accelerator groups.
\begin{itemize}
    \item \textbf{Intra-Group Homogeneous Communication:} Within a group of homogeneous accelerators (e.g., all NVIDIA GPUs or all Cambricon MLUs involved in a collective operation), KAITIAN directs communication tasks (e.g., AllReduce, Broadcast) to the vendor's highly optimized communication library. For instance, operations solely among GPUs would use NCCL, while operations solely among MLUs would use CNCL. This ensures near-native performance for communication within such homogeneous groups.
    \item \textbf{Inter-Group Heterogeneous Communication:} For communication between different types of accelerator groups (e.g., exchanging gradients between a GPU-based group and an MLU-based group), KAITIAN employs a general-purpose communication layer, Gloo. Data transfer between different accelerator types is managed by:
        \begin{enumerate}
            \item Copying the tensor from the source accelerator's memory to host (CPU) RAM.
            \item Transmitting the tensor data from the source host's RAM to the target host's RAM using Gloo's TCP/IP backend (or other suitable Gloo backends if available for host-to-host communication).
            \item Copying the tensor from the target host's RAM to the target accelerator's memory.
        \end{enumerate}
\end{itemize}
This explicit staging through host memory is necessary because direct memory-to-memory communication between, for example, an NVIDIA GPU and a Cambricon MLU is generally not supported at the hardware or driver level. While this relay introduces additional memory copy overhead (accelerator-to-host, host-to-accelerator), it provides a universal mechanism for interoperability. KAITIAN's design aims to minimize the impact of such inter-group transfers, typically limiting them to essential synchronization points like gradient aggregation across all involved devices.

\subsection{Hybrid Communication Backend Integration}
To implement this hierarchical strategy, KAITIAN extends PyTorch's `torch.distributed` C++ backend by introducing `ProcessGroupKaiTian`. This meta-process group can internally manage or dispatch to multiple underlying communication backends. When a collective operation is initiated by PyTorch DDP using `ProcessGroupKaiTian`:
\begin{enumerate}
    \item KAITIAN analyzes the participating processes and their device types to determine if the operation is within a single type of accelerator group (homogeneous) or spans multiple, different accelerator types (heterogeneous).
    \item If the operation is homogeneous (e.g., an AllReduce among a set of GPUs), it dispatches the call to the appropriate vendor library (e.g., NCCL) for that group.
    \item If the operation is heterogeneous (e.g., an AllReduce involving both GPUs and MLUs), it orchestrates the multi-step transfer via Gloo and host memory as described above.
\end{enumerate}
This allows a single PyTorch distributed training script to transparently utilize a mix of accelerators without requiring the user to manually manage the underlying communication complexities for different hardware.

\subsection{Load-Adaptive Scheduling Mechanism}
Heterogeneous accelerators invariably possess different raw processing capabilities and memory bandwidths. In synchronous distributed training (like data-parallel AllReduce SGD), the overall pace is dictated by the slowest worker (straggler). Without load balancing, faster devices would frequently idle waiting for slower ones, leading to poor resource utilization. KAITIAN’s load-adaptive mechanism aims to mitigate this by dynamically distributing the workload proportionally to each accelerator's effective processing speed.

The mechanism operates in two phases:
\begin{enumerate}
    \item \textbf{Offline Benchmarking (Optional) / Online Profiling:}
        \begin{itemize}
            \item \textbf{Initial Benchmarking:} Before the main training loop, KAITIAN can optionally run a short profiling job (e.g., a few forward/backward passes of the target model with a small, fixed amount of data) on each participating accelerator. This provides an initial estimate of their relative speeds. The fastest device is assigned a score of 1.0, and other devices $i$ are scored relative to it: $\text{score}_i = \text{time}_{\text{fastest}} / \text{time}_i$.
            \item \textbf{Online Adaptation (Future Work):} While the current implementation primarily uses initial benchmarking, a more advanced version could continuously monitor computation times per batch during training and dynamically adjust scores. This would account for variations in performance due to factors like thermal throttling or contention for shared resources.
        \end{itemize}
    \item \textbf{Dynamic Data Allocation:} During data loading for each training step, KAITIAN's custom `DistributedSampler` (or a similar data partitioning mechanism) allocates portions of the global mini-batch to each device $i$ based on its score. If the total global batch size is $B_{global}$, and there are $N$ devices, the batch size for device $i$, $b_i$, is calculated such that it is proportional to $\text{score}_i$, while ensuring $\sum_{i=1}^{N} b_i = B_{global}$. A common way to achieve this is:
        $$ \text{batch\_size}_i = \frac{\text{score}_i}{\sum_{j=1}^{N} \text{score}_j} \times B_{\text{global}} $$
        The allocated batch sizes are then rounded to the nearest integer, ensuring the sum matches $B_{\text{global}}$. This aims to equalize the computation time across all devices for each step, as $T_i = \text{Workload}_i / \text{Speed}_i \approx (k \cdot \text{batch\_size}_i) / (c \cdot \text{score}_i)$, where $k$ and $c$ are constants. By making $\text{batch\_size}_i \propto \text{score}_i$, $T_i$ should ideally be similar for all devices.
\end{enumerate}
This load-adaptive approach ensures that faster accelerators process more data, while slower ones handle a proportionally smaller load, leading to more balanced computation times per iteration and improved overall training throughput.

\subsection{System Coordination}
KAITIAN utilizes a lightweight coordination service, such as Redis, for initial process discovery, group membership management, and synchronization of metadata (e.g., benchmark scores, rendezvous information). This is a common pattern in distributed PyTorch setups.

By combining these design elements—hierarchical communication management via a dedicated ProcessGroup, hybrid backend integration, and load-adaptive scheduling—KAITIAN provides a robust and efficient solution for harnessing the power of heterogeneous accelerator clusters in demanding embodied AI applications.

\section{Implementation and Evaluation}
\label{sec:implementation_evaluation}
To validate the efficacy and performance of the KAITIAN framework, we implemented it as an extension to PyTorch's distributed communication library and conducted a series of experiments on a dedicated heterogeneous hardware testbed.

\subsection{Implementation Details}
KAITIAN's core logic is implemented in C++ as a custom process group backend for PyTorch (version 1.10 was used for development, but it's designed to be adaptable to newer versions). This `ProcessGroupKaiTian` backend interfaces with vendor libraries (NCCL 2.11, CNCL 1.5) and Gloo (PyTorch's bundled version).
\begin{itemize}
    \item \textbf{Process Group Management:} We extended PyTorch's `ProcessGroup.hpp` and related classes to create `ProcessGroupKaiTian`. This class is responsible for managing subgroups for homogeneous communication (e.g., by internally invoking `ProcessGroupNCCL` or `ProcessGroupCNCL` for operations confined to those device types) and orchestrates inter-group transfers using `ProcessGroupGloo`.
    \item \textbf{DistributedSampler Override:} For load-adaptive scheduling, we implemented a custom `KaitianDistributedSampler` that overrides PyTorch's default `DistributedSampler`. This sampler takes per-device scores as input and distributes dataset indices accordingly.
    \item \textbf{Control Plane:} A set of Python utility scripts and command-line tools are provided to launch and manage KAITIAN training jobs. These scripts handle the setup for different accelerator environments and use a Redis server for rank discovery, initial handshake, and sharing benchmark scores.
    \item \textbf{Environment Isolation in Experiments:} For experimental purposes and to ensure reproducibility across different hardware stacks, Docker containers were employed. These containers encapsulated the distinct software environments required for NVIDIA GPUs (CUDA Toolkit, NVIDIA drivers) and Cambricon MLUs (CNToolkit, Cambricon drivers), preventing library conflicts. This use of Docker is an implementation detail of the experimental setup, not a conceptual part of the KAITIAN framework's core communication logic itself.
\end{itemize}
The implementation effort focused on creating a clean interface that abstracts the underlying complexity from the end-user, allowing them to specify heterogeneous configurations with relative ease.
% \placeholder{Optionally, add a small code snippet showing how a user might initialize KAITIAN or define a heterogeneous group in their PyTorch script, if it's concise and illustrative.}

\subsection{Experimental Setup}
\label{ssec:experimental_setup}
\begin{itemize}
    \item \textbf{Hardware Testbed:}
        \begin{itemize}
            \item CPU: AMD EPYC 7763 (64-core)
            \item RAM: 64 GB DDR4
            \item GPUs: 2 x NVIDIA GeForce GTX 1080 (Pascal architecture, 8 GB VRAM each)
            \item NPUs: 2 x Cambricon MLU370-S4 (ShangNeng architecture, 16 GB VRAM each) % Please verify.
            \item Interconnect: PCIe Gen3 for accelerators. Gigabit Ethernet for host-level communication if Gloo uses TCP/IP across different physical nodes (though in this setup, all devices are in one server, so Gloo likely uses shared memory or local loopback for CPU-level transfers).
        \end{itemize}
    \item \textbf{Software Environment:}
        \begin{itemize}
            \item OS: Ubuntu 20.04 LTS
            \item Containerization (for experimental isolation): Docker 20.10
            \item NVIDIA Stack: CUDA Toolkit 11.2, NVIDIA Driver 460.xx, NCCL 2.11
            \item Cambricon Stack: CNToolkit 3.2.2, Cambricon Driver 5.9.4, CNCL 1.5 % Please verify these versions
            \item Deep Learning Framework: PyTorch 1.10 (compiled from source with custom backend)
            \item Coordination: Redis 6.0
        \end{itemize}
    \item \textbf{Benchmark Task:}
        \begin{itemize}
            \item Model: MobileNetV2~\cite{Sandler:MobileNetV2} (a widely used efficient CNN architecture, suitable for embodied AI).
            \item Dataset: CIFAR-10~\cite{Krizhevsky:CIFAR10} (32x32 color images, 10 classes). While small, it allows for rapid iteration and focuses evaluation on communication and scheduling overheads rather than being I/O bound by a massive dataset.
            \item Training Configuration:
                \begin{itemize}
                    \item Optimizer: Stochastic Gradient Descent (SGD) with momentum 0.9, weight decay 5e-4.
                    \item Learning Rate: Initial LR 0.1, with a step decay schedule.
                    \item Global Batch Size: 256 (distributed across devices).
                    \item Epochs: 50 (sufficient to observe training trends and convergence).
                    \item Loss Function: Cross-Entropy Loss.
                \end{itemize}
        \end{itemize}
    \item \textbf{Baselines and Configurations Evaluated:}
        \begin{itemize}
            \item \textbf{2G (NCCL)}: Homogeneous training using 2 NVIDIA GPUs with NCCL.
            \item \textbf{2M (CNCL)}: Homogeneous training using 2 Cambricon MLUs with CNCL.
            \item \textbf{KAITIAN (1G+1M)}: Heterogeneous training with 1 GPU and 1 MLU.
            \item \textbf{KAITIAN (2G+1M)}: Heterogeneous training with 2 GPUs and 1 MLU.
            \item \textbf{KAITIAN (1G+2M)}: Heterogeneous training with 1 GPU and 2 MLUs.
            \item \textbf{KAITIAN (2G+2M)}: Heterogeneous training with 2 GPUs and 2 MLUs.
        \end{itemize}
        For KAITIAN configurations, the load-adaptive mechanism was active.
\end{itemize}

\subsection{Evaluation Metrics}
The primary metrics for evaluation were:
\begin{itemize}
    \item \textbf{Total Training Time:} Wall-clock time to complete 50 epochs. This is the key indicator of end-to-end performance.
    \item \textbf{Model Accuracy:} Top-1 accuracy on the CIFAR-10 test set after 50 epochs, to ensure that KAITIAN does not negatively impact convergence.
    \item \textbf{Communication Overhead (Homogeneous):} Comparison of training time using KAITIAN with only homogeneous devices (e.g., 2 GPUs managed by KAITIAN but only using its NCCL path) versus native NCCL/CNCL, to quantify the overhead introduced by KAITIAN's framework layer.
    \item \textbf{Scalability and Resource Utilization:} Inferred from training time improvements as more heterogeneous devices are added and from the effectiveness of the load-adaptive mechanism.
\end{itemize}

\subsection{Results and Analysis}

\subsubsection{Training Efficiency and Accuracy}
\begin{figure}[htbp]
    \centering
    \includegraphics[width=\columnwidth]{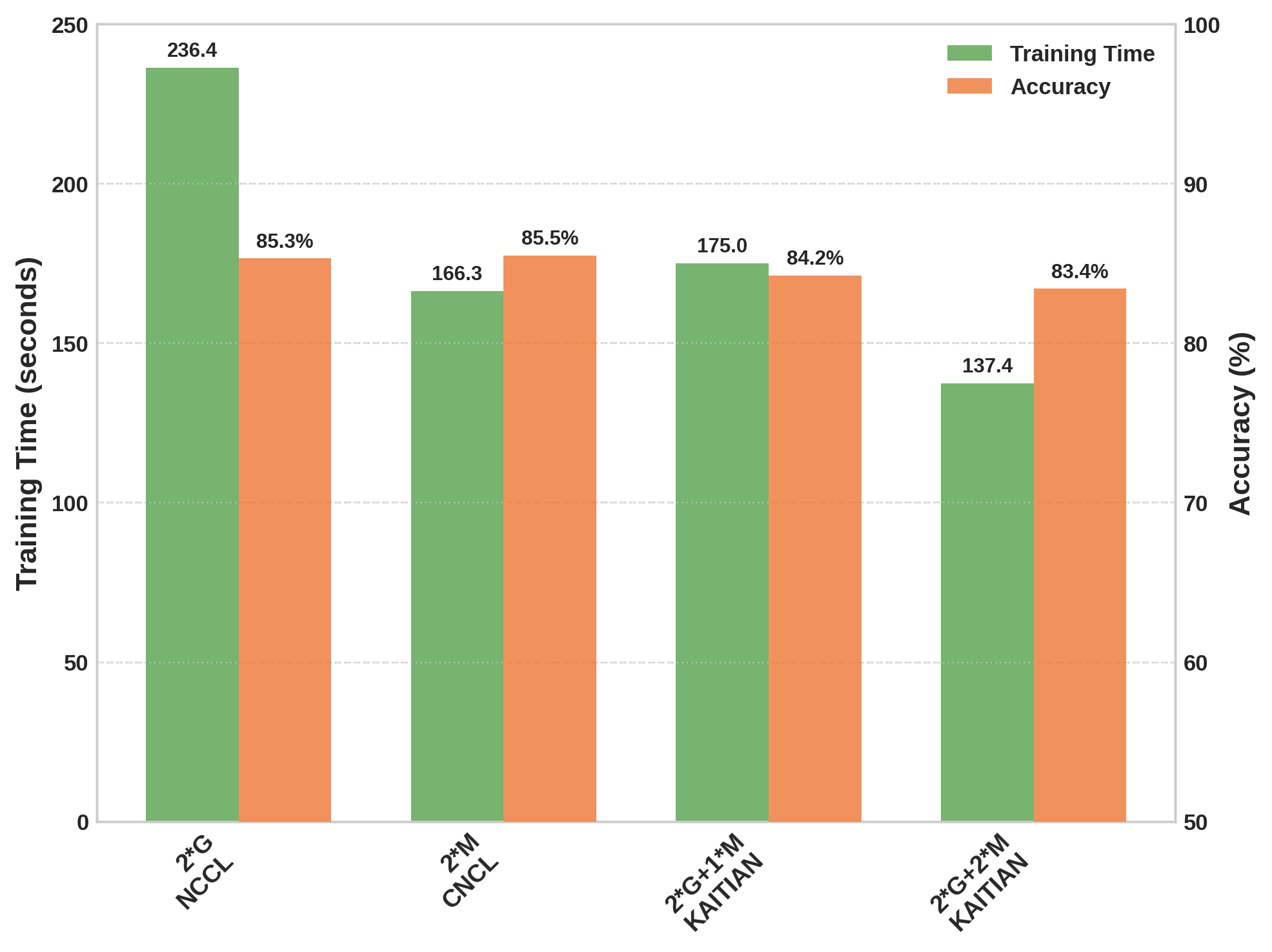} % This is the existing fig2.png for results
    \caption{KAITIAN Training Efficiency (Time to Complete 50 Epochs) and Model Accuracy Comparison on CIFAR-10 with MobileNetV2. Lower training time is better. Accuracy is Top-1 on the test set.}
    \label{fig:training_efficiency_main}
    \vspace{-1em}
\end{figure}

Figure~\ref{fig:training_efficiency_main} shows the total training time and final model accuracy for various configurations.

Key observations:
\begin{itemize}
    \item \textbf{Heterogeneous Speedup:} The KAITIAN (2G+2M) configuration achieved the fastest training time of 137.4 seconds. This represents a significant speedup of approximately 42\% compared to the 2G (NCCL) baseline (236.4 seconds) and about 17\% compared to the 2M (CNCL) baseline (166.3 seconds). This clearly demonstrates KAITIAN's ability to effectively harness the combined computational power of diverse accelerators.
    \item \textbf{Scalability:} As more accelerators were added in heterogeneous configurations (e.g., moving from 1G+1M to 2G+1M, then to 2G+2M), the training time generally decreased, indicating good scalability. For instance, 2G+1M (175.0s) was faster than both 2G (236.4s) and 1M (not shown, but would be slower than 2M's 166.3s / 2 = ~332s if linear).
    \item \textbf{Model Accuracy:} Accuracy remained comparable across all configurations. The 2G+2M KAITIAN setup achieved 83.4\% accuracy, while the 2G (NCCL) baseline reached 85.3\% and the 2M (CNCL) baseline reached 85.5\%. The slight variations (around 2\%) are within typical experimental noise for CIFAR-10 training runs and suggest that KAITIAN's communication and scheduling mechanisms do not adversely affect model convergence or final performance. The primary goal of such a framework is to accelerate training without degrading accuracy, which KAITIAN achieves.
\end{itemize}

\subsubsection{Effectiveness of Load-Adaptive Mechanism}
\begin{figure}[htbp]
    \centering
    \includegraphics[width=0.9\columnwidth]{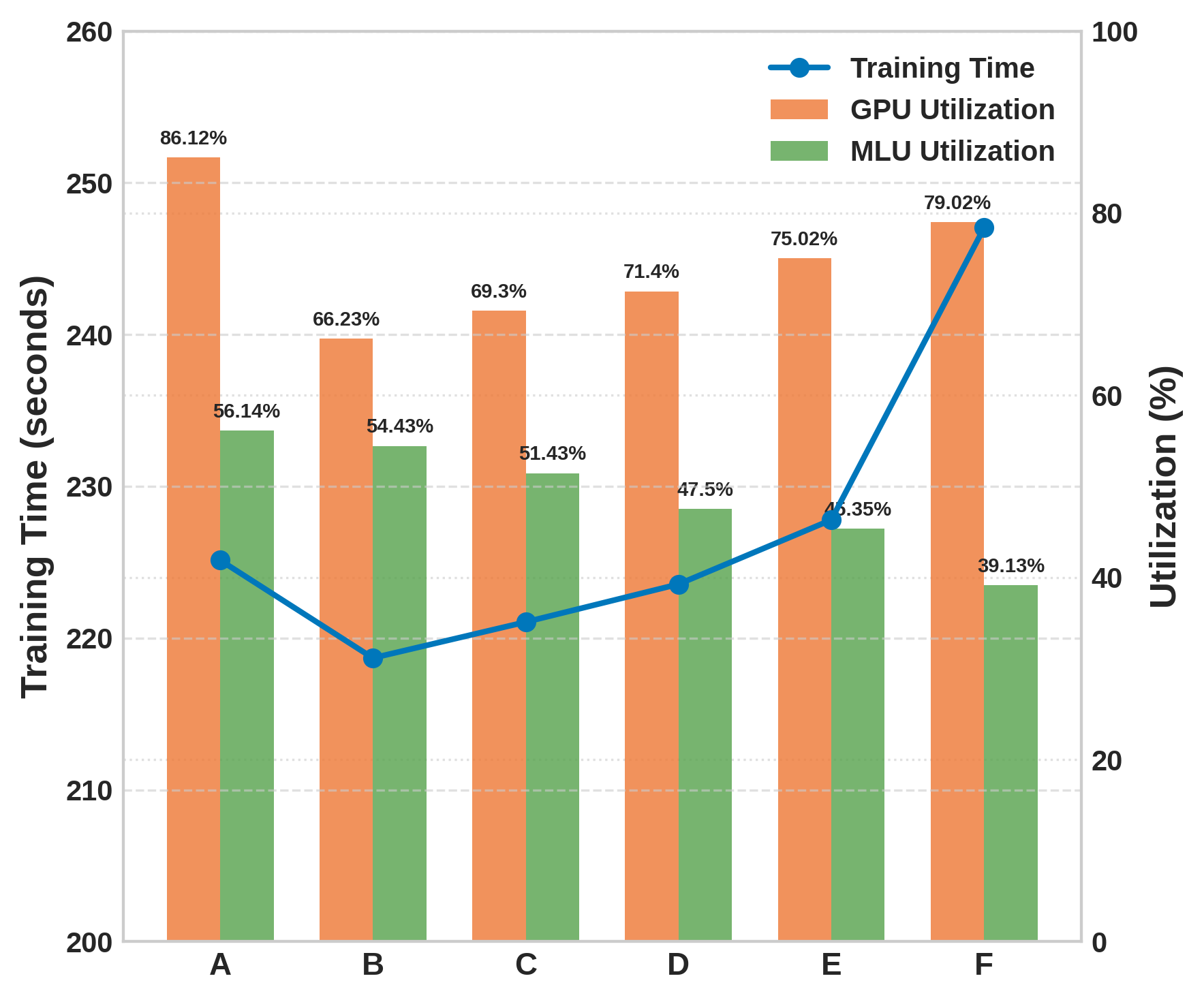} % This is the existing fig5.png for load adaptation
    \caption{Impact of Load Adaptive Mechanism on Heterogeneous Training (e.g., 1G+1M configuration). Strategy A might be naive equal batch splitting, Strategy B KAITIAN's adaptive splitting, Strategy C a suboptimal fixed ratio. The y-axis likely represents training time per epoch or overall, and x-axis different strategies or device utilization ratios. KAITIAN (Strategy B) finds a balance that minimizes training time by ensuring both GPU and MLU are effectively utilized according to their capabilities.}
    \label{fig:load_adaptive_effect_main}
    \vspace{-1em}
\end{figure}

Figure~\ref{fig:load_adaptive_effect_main} illustrates the impact of the load-adaptive mechanism.

To elaborate on the figure's implications:
\begin{itemize}
    \item Without load adaptation (e.g., if a naive 50/50 split of batches was used between a faster GPU and a slower MLU), the MLU would become a bottleneck, and the GPU would be underutilized. This would correspond to a suboptimal point (like "Strategy A" or "C" if they represent imbalanced fixed allocations).
    \item KAITIAN's mechanism, by benchmarking and assigning scores (e.g., GPU score 1.0, MLU score 0.7 if MLU is 70% as fast for the given task), allocates more data to the GPU. This ensures both devices finish their portion of the batch at roughly the same time, leading to "balanced GPU and MLU utilizations" and thus minimizing the per-iteration time and overall training time (represented by "Strategy B").
    \item This result underscores the critical importance of load balancing in heterogeneous environments. Even with an efficient communication layer, performance gains can be nullified if workloads are not intelligently distributed.
\end{itemize}
% \placeholder{If possible, quantify the performance difference between KAITIAN's adaptive strategy and a naive equal-split strategy for a specific heterogeneous setup, e.g., "Without load adaptation, the 1G+1M setup took X seconds, while with adaptation it took Y seconds, an improvement of Z\%."}

\subsubsection{Communication Overhead Analysis}
\begin{figure}[htbp]
    \centering
    \includegraphics[width=0.9\columnwidth]{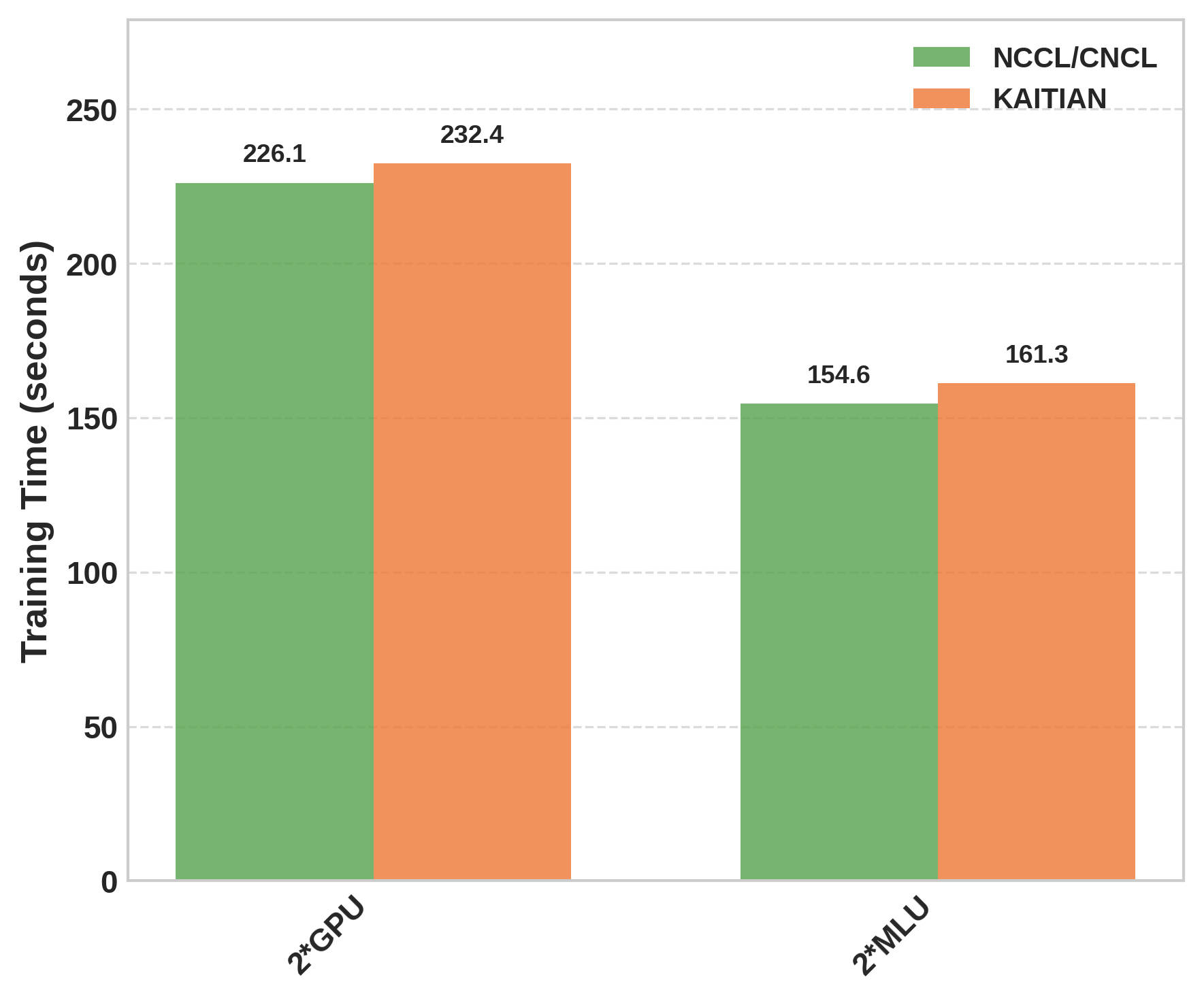} % This is the existing fig4.png for overhead
    \caption{Communication Overhead of KAITIAN in Homogeneous Settings. Comparison of training time using native NCCL/CNCL versus KAITIAN managing the same homogeneous devices.}
    \label{fig:comm_overhead_main}
    \vspace{-1em}
\end{figure}

Figure~\ref{fig:comm_overhead_main} quantifies the overhead introduced by KAITIAN's framework layer when operating in a purely homogeneous setting.

The results indicate:
\begin{itemize}
    \item \textbf{GPU Homogeneous:} For 2 GPUs, native NCCL training took 226.1 seconds. When KAITIAN managed these 2 GPUs (internally still using NCCL for the actual communication), the time was 232.4 seconds. This represents an overhead of (232.4 - 226.1) / 226.1 $\approx$ 2.8\%.
    \item \textbf{MLU Homogeneous:} For 2 MLUs, native CNCL training took 154.6 seconds. With KAITIAN managing these 2 MLUs (internally using CNCL), the time was 161.3 seconds. This is an overhead of (161.3 - 154.6) / 154.6 $\approx$ 4.3\%.
\end{itemize}
These overheads are minimal and demonstrate that KAITIAN's abstraction layer for managing multiple backends and its data routing logic impose only a slight performance penalty when operating on homogeneous devices. This is crucial, as it means users do not pay a significant "KAITIAN tax" for the added flexibility when heterogeneity is not actively being exploited for a particular sub-task. The overhead is likely due to the extra dispatch logic within KAITIAN's custom process group.

Overall, the experimental results validate that KAITIAN successfully enables efficient distributed training across heterogeneous accelerators, achieving substantial speedups by effectively utilizing combined resources and intelligently balancing loads, all while introducing minimal framework overhead and maintaining model accuracy.

\section{Discussion}
\label{sec:discussion}
The development and evaluation of KAITIAN offer valuable insights into the challenges and opportunities of heterogeneous computing for embodied AI. Our results demonstrate that by thoughtfully integrating existing communication technologies and implementing adaptive scheduling, significant performance gains can be unlocked.

\subsection{Implications for Embodied AI}
Embodied AI systems often operate under tight latency and energy constraints. The ability to efficiently utilize a mix of accelerators (e.g., power-efficient NPUs for routine inference, high-performance GPUs for bursts of complex computation or on-device training/adaptation) is critical. KAITIAN provides a foundational software layer that can facilitate such flexible hardware utilization. For instance:
\begin{itemize}
    \item \textbf{Faster Model Development Cycles:} Researchers can iterate more quickly by distributing training across all available accelerators, regardless of vendor.
    \item \textbf{Deployment of Larger Models:} By pooling resources, more complex and capable AI models could potentially be trained and even deployed on edge platforms that feature heterogeneous SoCs (System-on-Chips).
    \item \textbf{Energy Efficiency:} Future extensions could incorporate power consumption into the load-adaptive mechanism, optimizing for energy efficiency in addition to speed, which is paramount for battery-powered robotic systems.
\end{itemize}

\subsection{Analysis of Inter-Group Communication Overhead}
The primary overhead in KAITIAN's inter-group communication stems from the data relay through host CPU memory (Device $\rightarrow$ CPU $\rightarrow$ Gloo $\rightarrow$ CPU $\rightarrow$ Device). While our results show substantial net speedups, this indirect path is inherently slower than direct device-to-device communication (like NVLink for GPU-GPU or specific inter-chip links). The performance gain from adding heterogeneous accelerators must outweigh this overhead. KAITIAN's success in the 2G+2M configuration suggests that for the MobileNetV2/CIFAR-10 task, the computational power added by the MLUs (and vice-versa for GPUs) was significant enough to overcome this. For tasks with very frequent or very large inter-group synchronizations, this overhead might become more dominant.
% \placeholder{Consider adding a microbenchmark result here: e.g., ping-pong latency or effective bandwidth for GPU-MLU transfer via KAITIAN vs. GPU-GPU via NCCL and MLU-MLU via CNCL.}

\subsection{Limitations}
Despite its promising results, KAITIAN has limitations:
\begin{itemize}
    \item \textbf{CPU Bottleneck:} The reliance on CPU-mediated transfers for inter-group communication can become a bottleneck if the CPU's processing power or memory bandwidth is insufficient, especially with many heterogeneous devices or very frequent synchronization.
    \item \textbf{Scope of Heterogeneity:} The current implementation supports GPUs and MLUs. Extending it to other accelerator types like FPGAs or specialized ASICs would require developing or integrating corresponding low-level communication primitives and device management code.
    \item \textbf{Dynamic Task Graphs:} KAITIAN is primarily designed for data-parallel training with regular synchronization. More complex distributed patterns, such as pipeline parallelism or models with highly dynamic computational graphs, might require more sophisticated scheduling and communication strategies.
    \item \textbf{Fault Tolerance:} The current framework does not explicitly address fault tolerance, which can be important in larger distributed systems.
    \item \textbf{Online Profiling Granularity:} The load-adaptive mechanism currently relies on initial benchmarking. A more fine-grained online profiler that adapts to performance fluctuations during training could yield further benefits but adds complexity.
\end{itemize}

\subsection{Future Work}
KAITIAN opens several avenues for future research and development:
\begin{itemize}
    \item \textbf{Advanced Load Balancing:} Incorporate online monitoring of computation times and potentially power consumption to create a more dynamic and multi-objective load balancing scheduler. Explore predictive models for task completion times on different accelerators.
    \item \textbf{Direct Inter-Device Communication Exploration:} Investigate emerging technologies or research efforts that might enable more direct (or lower-overhead) communication paths between different vendors' accelerators, potentially bypassing the CPU for certain transfers (e.g., using technologies like CXL if supported by future devices).
    \item \textbf{Broader Accelerator Support:} Extend KAITIAN to support a wider range of accelerators, including FPGAs and other NPUs, by developing new device-specific interface modules for `ProcessGroupKaiTian`.
    \item \textbf{Support for Diverse Parallelism Strategies:} Enhance KAITIAN to support other forms of parallelism beyond data parallelism, such as model and pipeline parallelism, across heterogeneous devices.
    \item \textbf{Robotic Platform Integration and Benchmarking:} The ultimate goal is to deploy and evaluate KAITIAN on real-world robotic platforms. This involves porting to embedded heterogeneous SoCs (e.g., NVIDIA Jetson series, Qualcomm Robotics platforms) and benchmarking on representative embodied AI tasks like SLAM (Simultaneous Localization and Mapping), visual servoing, and grasp planning. This will expose new challenges related to resource constraints and real-time performance.
    \item \textbf{Integration with Higher-Level Workflow Managers:} Explore integration with workflow managers like Kubeflow or Ray to simplify the deployment and management of KAITIAN-accelerated applications in larger clusters.
\end{itemize}

By addressing these areas, KAITIAN can evolve into an even more powerful and versatile framework for the next generation of embodied AI systems.

\section{Conclusion}
\label{sec:conclusion}
In this paper, we introduced KAITIAN, a novel distributed communication framework designed to overcome interoperability barriers and unlock the potential of heterogeneous accelerator clusters for embodied AI applications. By strategically integrating vendor-specific communication libraries for high-performance intra-group communication with a general-purpose layer for inter-group data exchange, and by incorporating a load-adaptive scheduling mechanism, KAITIAN enables efficient collaboration between diverse hardware like NVIDIA GPUs and Cambricon MLUs.

Our implementation within PyTorch and comprehensive evaluations on an image classification task demonstrated that KAITIAN can significantly accelerate distributed training—achieving up to a 42\% reduction in training time in a 2-GPU + 2-MLU configuration compared to homogeneous baselines. This performance gain is achieved with minimal communication overhead (2.8--4.3\% in homogeneous settings) and without compromising model accuracy. The load-adaptive mechanism proved crucial in balancing workloads and maximizing resource utilization across devices with varying computational capabilities.

KAITIAN represents a significant step towards more flexible, scalable, and powerful computing paradigms for embodied AI. It empowers researchers and developers to leverage the full spectrum of available hardware resources, accelerating the development and deployment of complex AI models that are vital for intelligent autonomous systems. Future work will focus on extending KAITIAN's capabilities, supporting a broader range of accelerators and parallelism strategies, and deploying it on physical robotic platforms to tackle real-world embodied intelligence tasks.

\section*{Acknowledgment}
This research was partially supported by the National Key Research and Development Program of China (2022YFB4501600).

\ifCLASSOPTIONcaptionsoff
  \newpage
\fi

% If you have biographies
% \begin{IEEEbiography}{Jieke Lin}
% Biography text here.
% \end{IEEEbiography}

% \begin{IEEEbiography}{Wanyu Wang}
% Biography text here.
% \end{IEEEbiography}

% \begin{IEEEbiography}{Longxiang Yin}
% Biography text here.
% \end{IEEEbiography}

% \begin{IEEEbiography}{Yinhe Han}
% Biography text here.
% \end{IEEEbiography}

\end{document}